\documentstyle[aps,preprint,prb]{revtex}
\begin{document}
\title{Variable curvature slab molecular dynamics 
as a method to determine surface stress}
\author{Daniele Passerone $^{a,b}$, Erio Tosatti$^{a,b,c}$, 
Guido L. Chiarotti$^{a,b}$ and Furio Ercolessi$^{a,b}$}
\address{$^{a)}$ International School for Advanced Studies (SISSA), 
Via Beirut 4, I-34014 Trieste, Italy\\
$^{b)}$ Istituto Nazionale di Fisica della Materia (INFM), 
Unit\`a Trieste SISSA, Italy\\
$^{c)}$ The Abdus Salam International Centre for Theoretical Physics, 
P.O.~Box 586, I-34014 Trieste, Italy}
\date{11 June 1998}
\maketitle

\begin{abstract}
A thin plate or slab, prepared so that opposite faces have different surface
stresses, will bend as a result of the stress difference. We have developed
a classical molecular dynamics (MD) formulation where (similar in spirit to
constant-pressure MD) the curvature of the slab enters as an additional
dynamical degree of freedom. The equations of motion of the atoms have been
modified according to a variable metric, and an additional equation of
motion for the curvature is introduced. We demonstrate the method to Au
surfaces, both clean and covered with Pb adsorbates, using many-body glue
potentials. Applications to stepped surfaces, deconstruction and other
surface phenomena are under study.
\end{abstract}

\vspace{1cm}
\begin{center}
Preprint SISSA {\bf 60/98/CM/SS}\\
Submitted to {\sl Physical Review B}
\end{center}
\clearpage
\tightenlines

\section{Introduction}

There has been recently a reevaluation of the role of surface stress, as an
important microscopic indicator of the state of a crystal surface. Recent
experimental and theoretical aspects of issues related to surface stress
have been reviewed by Ibach \cite{ibach}.

A good example of connection between micro- and macroscopic quantities is
the determination of surface stress from the measurement of macroscopic
deformations of a thin plate-shaped sample. According to elasticity theory 
\cite{landau}, if a thin plate presents a stress difference between its two
surfaces, it will bend in order to minimize its total free energy. In the
limit of small deformations, a simple formula connects stress difference,
thickness, curvature and elastic properties of the sample. Several methods
have been recently used to measure the bending. They are mainly optical or
based on scanning tunneling microscopy (STM)\cite{ibach}. 
Flinn, Gardner and Nix \cite{flinn}
discussed a laser scanning technique to measure the stress-induced
curvature, and presented experimental results for stress in Al-Si films as a
function of temperature. Martinez, Augustyniak and Golovchenko \cite
{golovchenko} measured surface-stress changes resulting from monolayer and
sub-monolayer coverages of gallium on Si$(111)$. They used LDA calculations
of stress in the Ga-covered Si$(111)$ for the first determination of stress
in the Si$(111)$ $7\times 7$ and Si(Ga) superlattice surfaces. Stress variations
associated with deconstruction of Au$(111)$ and Au$(100)$ were obtained by
STM \cite{bach1,bach2}. Moreover, the influence of adsorbates on surface
stress has been exploited using the sample bending method for C/Ni(111) \cite
{ads1}, S/Ni(111) \cite{ads2}, Co/Ni(100) \cite{ads3}, K/Pt(111) \cite{ads4}%
, Co/Pt(111) \cite{ads5}, Ag/Pt(111) \cite{ads6}.

An atomistic simulation method which controls the curvature of a sample as
an extra degree of freedom has not been available so far, although this
would seem a potentially useful tool. The situations in which surface stress
varies when physical phenomena occur at interfaces are numerous: adsorbates,
reconstructions, steps, phase transitions can change surface stress
dramatically. All of these phenomena are, under some conditions, well
described by suitable MD simulations. 
A simulation method where the slab curvature can be measured
as a function of adsorbate coverage
should represent a powerful connection between important microscopic and
macroscopic quantities.

In this work we describe and demonstrate a scheme which is meant to fill
this gap. It is a classical MD method, but it could in principle be
extended to {\em ab initio} MD calculations. It is based on extending the usual
concept of variable cell MD, pionereed by Andersen \cite{andersen} and by
Parrinello and Rahman \cite{parr_rahm}, to a slab with variable curvature.
In our formulation the curvature is a single, global Lagrangian degree of
freedom. Surface stress can be extracted as a direct result of the
calculation through elasticity equations \cite{stoney}, which will be
described below.

The outline of the work is as follows: section \ref{sec2} will present the
theory and geometric considerations underlying our simulation. In \ref{sec3}
the phenomenology of a bent plate is reviewed in order to extract the
pertinent equations. In section \ref{sec4} we present some initial
applications of the method. Finally, section \ref{sec5} is devoted to
discussion and conclusions.

\section{Theory}

\label{sec2}

In ordinary variable-cell MD, the coordinates of an atom can be written as 
\begin{equation}
{\bf r}_i{\bf =\hat{H}\cdot {s}}_i,
\end{equation}
where ${{\bf r}_i}$ is the position vector of the $i$-th particle, ${{\bf s}%
_i}$ are scaled coordinates ($s_i^j,j=1\dots 3$ are held between $-0.5$ and 
$0.5$) and ${\bf \hat{H}}$ is a matrix describing the space metric inside the
cell. Supposing, as was done by Parrinello and Rahman \cite{parr_rahm}, that
the cell can vary in volume and shape, these variations can be accounted for
by ${\bf \hat{H}}$, whose elements can be treated as an extra set of
dynamical degrees of freedom. Our aim is to extend this kind of approach to
the case of a bending plate, through a different choice of the metric ${\bf 
\hat{H}}$.

Let us start with a slab-shaped system, with two surfaces and lateral $%
(x,y) $ periodic boundary conditions (PBC). Suppose we bend the slab
cylindrically through a radius $R$ (see Figure 1). A convenient choice of
the transformation matrix, convenient to our problem, could be the following:

\begin{equation}
{\bf {r}=\hat{H}}\left( {\bf {s}},R\right) \cdot {{\bf s}\,,}
\label{eq:r_and_s}
\end{equation}
where ${\bf \hat{H}}$, depending parametrically on the radius of curvature $%
R$, is no longer uniform (as in the Parrinello-Rahman scheme), but
rather depends on the point ${\bf {s}}$. This dependence is what originates
the curvature. In Figure 1, for example, $s_1$ goes from $-0.5$ to $0.5$ 
along $x$, which we choose to be unaffected by curvature, $s_2$ goes from 
$-0.5$ to $0.5$ along direction $y$ which follows the curvature, and $s_3$ 
spans the sample from its inner to its outer surface.

From now on, we will use as reference surface an ideal ``neutral cylinder''
inside the slab, defined by $s_3=0$. In particular, we define the
curvature $k$ as the inverse radius of the neutral cylinder:
\begin{equation}
k=\left. \frac 1R\right| _{s_3=0}.
\end{equation}
In this way, if $L_y$ is the linear dimension (arc) along the bending
direction at $k=0$, the total bending angle is given by
$\theta _{M}=L_y\,k$.
Whereas in the Parrinello-Rahman approach the whole
matrix ${\bf \hat{H}}$ is treated as a set of additional degrees of freedom,
in our case ${\bf \hat{H}}$ is completely determined by the single parameter 
$k$ which we will include in the Lagrangian as an extra degree of freedom.

\subsection{Lagrangian}

The next step is the construction of a Lagrangian. The general, classical
Lagrangian for a $N$-particle interacting system is: 
\begin{equation}
{\em L}={\em T-V}=\frac 12\sum_im_i\left\| {\bf \dot {r}}_i\right\| ^2-{\em V%
}{\cal \,}\left( {\bf r}_1\dots {\bf r}_N\right)
\end{equation}
In Cartesian coordinates ${\bf \dot{r}}_i{\bf =\hat{H}\cdot \dot{s}}_i$ and $%
\left\| {\bf \dot{r}}\right\| ^2={\bf \dot{s}}_i^{{\bf T}}{\bf \hat{G}\,\dot{%
s}}_i$, where ${\bf \hat{G}=\hat{H}}^{{\bf T}}{\bf \hat{H}}$ is the metric
tensor\cite{parr_rahm}.
We now define an extended Lagrangian ${\em \tilde{L}}$ by including an
artificial ``curvature kinetic energy'':
\begin{equation}
{\em \tilde{L}\,}=\,{\em L}+\frac 12W\,\dot{k}^2,
\end{equation}
where $W$ acts effectively as a ''curvature inertial mass''. The velocity of
a particle can be written as: 
\begin{equation}
{\bf \dot{r}=}\frac{{\rm d}}{{\rm dt}}\left( {\bf \hat{H}\cdot s}\right) ,
\end{equation}
or explicitly 
\begin{equation}
\dot{r}^\alpha =\frac{{\rm d}}{{\rm dt}}\left( H_{\alpha \beta }\,s^\beta
\right) =\dot{H}_{\alpha \beta }\,\left( {\bf s\,,}k\right) \,s^\beta
+H_{\alpha \beta }\,\dot{s}^\beta =\left[ \frac{\partial H_{\alpha \beta }}{%
\partial s^\gamma }\dot{s}^\gamma +\frac{\partial H_{\alpha \beta }}{%
\partial k}\dot{k}\right] \,s^\beta +H_{\alpha \beta }\,\dot{s}^\beta .
\label{eq:time_derivative}
\end{equation}
The third term is the usual time derivative for the constant-cell motion,
whereas the second term contains a time derivative of 
the extra degree of freedom $k$. As in the Parrinello-Rahman 
formulation \cite{parr_rahm}, we will omit this term from the
definition of the particle velocities.
The kinetic energy of the system will therefore depend on $\dot{k}$
only through the curvature kinetic energy,
\begin{equation}
{\em \tilde{T}}{\cal =}\frac 12W\,\dot{k}^2,
\end{equation}
whose interpretation is discussed in Appendix A.
Parrinello and Rahman \cite{parr_rahm} and Andersen \cite{andersen}, using a
similar approach, showed that the equations of motion derived from their
Lagrangian satisfy important requisites, leading in particular to correct
thermodynamic averages (not dependent on the choice of $W$),
and to a correct balance between external and internal stress.

One can then cast (\ref{eq:time_derivative}) as
\begin{equation}
\dot{r}^\alpha =\frac{\partial H_{\alpha \beta }}{\partial s^\gamma }\dot{s}%
^\gamma \,s^\beta +H_{\alpha \beta }\,\dot{s}^\beta =\left[ \frac{\partial
H_{\alpha \sigma }}{\partial s^\beta }s^\sigma +H_{\alpha \beta }\right] \,%
\dot{s}^\beta \doteq M_{\alpha \beta }\,\dot{s}^\beta .
\end{equation}
The kinetic energy of a particle then becomes
\begin{equation}
{\em T}_1=\frac 12m\,\left\| {\bf \dot{r}}\right\| ^2=\frac 12m\,\dot{s}%
^\alpha N_{\alpha \beta }\,\dot{s}^\beta =
\frac 12m\, {\bf \dot{s}}^{{\bf T}}{\bf \hat{N}\,\dot{s}} \,\, ,
\label{eq:T1}
\end{equation}
where ${\bf \hat{N}\doteq \hat{M}}^{{\bf T}}{\bf \cdot \hat{M}.}$ 

We need now to express the potential energy in terms of the scaled coordinates
and of the metric. In the following, we will use mostly potentials (such as 
pairwise potentials or many-body potentials of the `glue' form) which only 
depend on pair distances: 
\begin{equation}
{\em V\,{\cal =}V\,}\left( \left\{ \left\| {\bf r}_i-{\bf r}_j\right\|
^2\left| i\ne j\,;i,j=1\dots N\right. \right\} \right) .
\end{equation}
It is therefore necessary to express $\left\| {\bf r}_i-{\bf r}_j\right\|
^2$ in the new coordinates \cite{extens}.
Denoting the metric tensor as ${\bf \hat{G}\,}\left( k,{\bf s}_i\right) =%
{\bf \hat{H}}^{{\bf T}}\left( k,{\bf s}_i\right) \cdot {\bf \hat{H}\,}\left(
k,{\bf s}_i\right) $ we have
\begin{equation}
\left\| {\bf r}_i-{\bf r}_j\right\| ^2\equiv r_{ij}^2={\bf s}_i^{{\bf T}}%
{\bf \hat{G}}\left( k,{\bf s}_i\right) \,{\bf s}_i+{\bf s}_j^{{\bf T}}{\bf 
\hat{G}}\left( k,{\bf s}_j\right) \,{\bf s}_j-2\,{\bf s}_i^{{\bf T}}\,{\bf 
\hat{H}\,}^{{\bf T}}\left( k,{\bf s}_i\right) \cdot {\bf \hat{H}\,}\left( k,%
{\bf s}_j\right) \,{\bf s}_j.
\end{equation}

We now have all the ingredients needed to write the $3N+1$ Lagrange
equations
\begin{equation}
\left\{ 
\begin{array}{l}
W\,\ddot{k}+\frac{\partial \,{\em V}}{\partial \,k}-\frac 12\sum_im_i\,{\bf 
\dot{s}}_i^{{\bf T}}\frac{\partial \,{\bf \hat{N}}}{\partial \,k}{\bf \dot{s}%
}_i=0 \\ 
\\ 
m_l{\bf \hat{N}}\left( k,{\bf s}_l\right) \,{\bf \ddot{s}}_l+\frac{\partial
\,{\em V}}{\partial \,{\bf s}_l}-\frac 12m_l\,{\bf \dot{s}}_l^{{\bf T}}
\frac{\partial\,{\bf \hat{N}}}{\partial \,{\bf s}_l}{\bf \dot{s}}_l+
m_l{\bf \dot{\hat{N}}}\left( k,\,{\bf %
s}_l\right) \,{\bf \dot{s}}_l=0,\,\,\,\,l=1\dots N.
\end{array}
\right.  \label{eq:motion}
\end{equation}

\subsection{Construction of the ${\bf \hat{H}}$ matrix}

To obtain an explicit form for (\ref{eq:r_and_s}) and (\ref{eq:motion}) we
write the expression for the Cartesian coordinates in a box with linear
dimensions (at $k=0$) $L_x$, $L_y$ and $L_z$:
\begin{equation}
\left\{ 
\begin{array}{l}
x=s_1L_x \\ 
y=\left( \frac 1k+s_3L_z\right) \sin \theta \\ 
z=s_3L_z\cos \theta +\frac 1k\left( \cos \theta -1\right) .
\end{array}
\right.  \label{eq:coor_tran}
\end{equation}
where $\theta $ is the angle running
along the $y$ direction, $-\theta_{M}/2<\theta<\theta_{M}/2$ (see Figure 1).
With the help of simple trigonometric formul\ae{}, we can 
express $\theta$ in terms of $s_2$ and $k$:
\begin{equation}
\sin \frac \theta 2=2\,s_2\sin \frac{k\,L_y}4
\end{equation}
whence equations (\ref{eq:coor_tran}) become:
\begin{equation}
\left\{ 
\begin{array}{l}
x=s_1L_x \\ 
y=4\,s_2\left( \frac 1k+s_3L_z\right) 
\sin \left( k\,L_y/4 \right)
\sqrt{1-4\left( s_2\right) ^2\sin ^2
\left( k\,L_y/4 \right) }\\ 
z=s_3L_z\left( 1-8\,\left( s_2\right) ^2\sin ^2\left( k\,L_y/4\right)
\right) -\frac 1k\left( 8\,\left( s_2\right) ^2\sin ^2\left( k\,L_y/4
\right) \right)
\end{array}
\right.
\end{equation}
By inspection, we see that a possible choice for ${\bf \hat{H}}$ is:
\begin{equation}
{\bf \hat{H}}\left( s_2,k\right) =\left( 
\begin{array}{ccc}
L_x & 0 & 0 \\ 
0 & \,\,\,\,\,\,4\frac{\sin (k\,L_y/4)}k
\sqrt{1-4\left( s_2\right) ^2\sin ^2 \frac{k\,L_y}4 %
}\,\,\,\,\, & \,\,\,\,\,4\,s_2L_z\sin \frac
{k\,L_y}4\sqrt{1-4\left( s_2\right) ^2\sin ^2\frac{k\,L_y}4} \\ 
0 & -s_2\frac{\sin ^2\left( k\,L_y/4\right) }k & L_z\left( 1-8\,\left(
s_2\right) ^2\sin ^2\left( \frac{k\,L_y}4\right) \right)
\end{array}
\right)  \label{eq:H_matrix}
\end{equation}

It can be easily shown that eq.~(\ref{eq:r_and_s}) is verified and that, in
the limit of zero curvature, $\lim_{k\rightarrow 0}{\bf \hat{H}}\left(
\,s_2,k\right) =%
\mathop{\rm diag}
\left[ L_x,L_y,L_z\right] $, that is, the Cartesian ${\bf \hat{H}}$ matrix is
recovered. Moreover, the same equations are exact in the limit of finite,
fixed curvature.

The explicit form of ${\bf \hat{N}}$ can be obtained by using equations (\ref
{eq:T1}) and (\ref{eq:H_matrix}): 
\begin{equation}
N_{\alpha \beta }{\bf =}\left( 
\begin{array}{ccc}
L_x^2 & 0 & 0 \\ 
0 & \frac{16\left( k\,L_z\,s_3+1\right) ^2\sin ^2\left( k\,u\right) }{\left(
1-4\left( s_2\right) ^2\sin ^2\left( k\,u\right) \right) k^2} & 0 \\ 
0 & 0 & L_z^2
\end{array}
\right) ,
\end{equation}
with $u=L_y/4$. ${\bf \hat{N}}$ is diagonal as a consequence of the
chosen cylindrical geometry.

\subsection{Constrained molecular dynamics}

The Lagrange equations are integrated using numerical methods. In
particular, we will use a second order velocity Verlet algorithm, which is
at the same time well-tested and simple. We found some problems due to the
instability of the center of mass of the sample during the bending, which
were overcome and will be commented just below.

Our starting geometry is an $n-$layer crystalline slab with PBC along the
$x$ and $y$ directions. The periodicity is accounted for by the scaled
coordinates $s_1$ and $s_2$. Along $s_1$, usual Cartesian PBC are adopted.
The boundary conditions for $s_2$ are also periodic, but the images of a
point of coordinates $(\bar{s}_1,\bar{s}_2,\bar{s}_3)$ have coordinates $(%
\bar{s}_1,\bar{s}_2\pm 1,\bar{s}_3)$, and lie 
on the cylindrical surface $s_3=\bar{s}_3$.
Every point of the sample is equivalent along $s_2$ and so
a ``translational'' invariance is still present,
albeit in a curved geometry.

For $k\neq 0$ the dynamics generated by (\ref{eq:motion}) does not conserve
the position of the center of mass of the slab along the radial
direction $s_3$.
To keep the center of mass fixed, we
resort to a constrained MD scheme. Constrained MD is a well known technique,
and we adopted the method used by Ryck\ae rt, Ciccotti and Berendsen \cite
{rickaert} and Andersen \cite{anderscons} for simulations of rigid
molecules. The main point to be kept in mind is that constraints must be
verified {\it exactly }at every time step, otherwise instabilities will
occur. Technical details on the practical implementation of this 
constraint are given in Appendix B.

The general MD protocol we apply consists of either constant energy
runs or constant
temperature runs (up to $300$ {\rm K}) obtained by velocity rescaling,
followed by quenching to $T=0$.
Occasionally, we also introduced a damping term in
the curvature dynamics. We chose a typical time step of the order of $%
10^{-14}{\rm s.}$

\section{Phenomenology of a bent plate}

\label{sec3}

Before moving on to the actual implementation, it is useful to recall some
known results concerning the phenomenology, largely contained in the review
by Ibach \cite{ibach}.

The definition of surface stress, according to Gibbs \cite{gibbs}, as the
``reversible work per unit area required to stretch a surface elastically'',
points out the difference of this physical quantity with respect to the
surface free energy, defined as the ``reversible work per unit area to
create a surface''. Stretching a {\it solid} surface implies modifying the
substrate, whereas a simple increase of area does not: whence the difference.

Let us start with the bulk stress tensor, whose element $\tau _{\,ij}$ is
defined as the $i$-th component of the force per unit area acting on the
side (with the normal to the surface parallel to the $j$-th direction) of a
small cube in the sample. The corresponding surface stress tensor can be
defined as the deviation of this quantity with respect to the bulk value,
integrated along the surface normal: 
\begin{equation}
\tau _{ij}^{\left( s\right) }=\int\nolimits_{-\infty }^{+\infty }{\rm dz\,}%
\left[ \tau _{ij}\left( \,z\right) -\tau _{ij}^{\left( b\right) }\right]
\end{equation}

The integral gives non-zero contribution only where the value of the stress
deviates from the bulk value, $i.e.$ at the surface.

Let us consider a rectangular shaped sheet of thickness $t$, delimited by
two identical $(100)$ surfaces. For our purposes the following simplified
construction can be put to use without loss of generality \cite{flinn}.
Schematize the plate at zero curvature as a ''substrate'' of thickness $t$
and length $L$ along $x_1,$ plus two films adsorbed on it: film $A$ on the
lower part of the plate, and film $B$ on the upper part. Let $L_A$ the
unconstrained length (along $x_1$) of film $A,$ and $L_B$ the length of film 
$B.$ Let $L_B>L$ and $L_A<L.$ In order to match the length $L$ of the
substrate, the film must be stretched (compressed) by an amount $\Delta
L_I=L-L_I,$ with $I=A,B$. The deformed film attached to an undeformed
substrate does not represent a condition of minimum free energy;\ the free
energy of the entire system can be lowered by deforming the substrate
slightly so as to reduce the deformation of the film. This deformation has
two components: an overall compression of the substrate and a bending of the
substrate.

It can be shown \cite{flinn} that the neutral plane ({\it i.e. }the plane
along which the strain is zero) can be set approximately in the middle of
the sheet, and the deviations from this approximation are negligible for
most cases. In the following, we will extract the conditions for the minimum
free energy of the sheet, allowed to bend along a single direction. For
simplicity, we will choose this to be the $[100]$ direction (of course
similar formul\ae{} can be derived for any other choice). If we denote the
coordinates along the bending direction, the non-bending direction and the
normal to the surface with $s_1,s_2$ and $s_3$ respectively, the situation
is similar to the one depicted in Figure 1. Each element of the sample is
subject to a strain $\varepsilon _{11}\left( \,x_3\right) $. The change in
free energy per unit area associated to this strain, of an element at height 
$s_3$ is: 
\begin{equation}
u\,\left( \,s_3\right) =\int\nolimits_0^{\varepsilon _{11}\left(
\,s_3\right) }\tau _{11}\left( \,s_3\right) \,{\rm d\,\varepsilon }_{11}.
\end{equation}
The total free energy change of the bent sample (per unit area) is:
\begin{equation}
U=\int\nolimits_{-\frac t2}^{+\frac t2}{\rm d}s_3\int\nolimits_0^{%
\varepsilon _{11}\left( \,s_3\right) }\tau _{11}\left( \,s_3\right) \,{\rm %
d\,\varepsilon }_{11}.
\end{equation}
We can separate this integral into two surface parts and a bulk part:
\begin{mathletters}
\label{eq:utot}
\begin{eqnarray}
U &=&U^{\,\left( s+\right) }+U^{\,\left( s-\right) }+U^{\,\left( b\right) }
\label{eq:utota} \\
U^{\,\left( s+\right) } &=&\int\nolimits_{+\frac t2-\alpha }^{+\frac t2}{\rm %
d}s_3\int\nolimits_0^{\varepsilon _{11}\left( \,s_3\right) }\left[ \tau
_{11}\left( \,s_3\right) -\tau _{11}^{\left( b\right) }\right] {\rm %
d\,\varepsilon }_{11}  \label{eq:utotb} \\
U^{\,\left( s-\right) } &=&\int\nolimits_{-\frac t2}^{-\frac t2+\beta }{\rm %
d}s_3\int\nolimits_0^{\varepsilon _{11}\left( \,s_3\right) }\left[ \tau
_{11}\left( \,s_3\right) -\tau _{11}^{\left( b\right) }\right] {\rm %
d\,\varepsilon }_{11}  \label{eq:utotc} \\
U^{\left( b\right) } &=&\int\nolimits_{-\frac t2}^{+\frac t2}{\rm d}%
s_3\int\nolimits_0^{\varepsilon _{11}\left( \,s_3\right) }\tau _{11}^{\left(
b\right) }\left( \,s_3\right) \,{\rm d\,\varepsilon }_{11}.  \label{eq:utotd}
\end{eqnarray}
The first two terms lead to the definition of surface stress in the Gibbs
sense, and can be written with good approximation:
\end{mathletters}
\begin{equation}
U^{\,\left( s+\right) }+U^{\,\left( s-\right) }=\left( \tau _{11}^{\left(
s\right) }\left( +\right) -\tau _{11}^{\left( s\right) }\left( -\right)
\right) \left( \frac{k\,t}2\right) ,
\end{equation}
where $\pm k\,t/2$ are the strains at the upper and lower surfaces.

The bulk term can be expressed using the elastic constants of the crystal.
Hooke's law relates strain and bulk stress tensor components in the
following way:
\begin{eqnarray}
\varepsilon _{11} &=&\sigma _{11}\,\tau _{11}+\sigma _{12}\,\tau _{22} \\
\varepsilon _{22} &=&\sigma _{12}\,\tau _{11}+\sigma _{11}\,\tau _{22}. 
\nonumber
\end{eqnarray}
In our case, $\varepsilon _{22}=0,$ therefore, after introducing Young's
modulus $Y$ and Poisson's ratio $\nu$:
\begin{eqnarray}
Y &=&\frac 1{\sigma _{11}}  \label{eq:Young} \\
\nu &=&-\frac{\sigma _{12}}{\sigma _{11}},  \nonumber
\end{eqnarray}
one obtains \cite{ibach}:
\begin{equation}
\varepsilon _{11}=\frac{1-\nu ^2}Y\tau _{11}.  \label{eq:epstau}
\end{equation}
If we substitute (\ref{eq:epstau}) into (\ref{eq:utotd}) we obtain, after
integration,
\begin{equation}
U^{\,\left( b\right) }=\frac 1{24}\frac Y{1-\nu ^2}\,k^2\,t^3,
\end{equation}
the total free energy of the bent plate is then:
\begin{equation}
F=\left( \tau _{11}^{\left( s\right) }\left( +\right) -\tau _{11}^{\left(
s\right) }\left( -\right) \right) \left( \frac{k\,t}2\right) +\frac 1{24}%
\frac Y{1-\nu ^2}\,k^2\,t^3.
\end{equation}
Minimizing $F$ with respect to the curvature, one obtains the equation
relating the equilibrium curvature $k_M$ to the difference in surface stress
between upper and lower surface:

\begin{equation}
\Delta \tau ^{\left( s\right) }=-\frac Y{6\,\left( 1-\nu ^2\right) }k_M\,t^2
\label{eq:stoney}
\end{equation}
where $Y$ is the Young's modulus, $\nu $ the Poisson's number, $k_M$ the
curvature and $t$ the thickness of the sample.
This relation is known as Stoney's equation, and was first derived in 1909
(except for the biaxial nature of the stress \cite{stoney,ibach}). 
We stress that
this formula is valid for a uniaxial bending, whereas for an isotropic
bending in two directions the factor $\left( 1-\nu ^2\right) $ should be
replaced by $\left( 1-\nu \right)$.
Another limit of validity is that the plate must be infinitely large,
relative to its thickness. In other words, there will be finite thickness
corrections in $(t/L)$ where $L$ is the lateral $(x,y)$ size. We did not
make any attempt to find these corrections so far.

In the remainder of this section we present the simple derivation of the
elastic constants two high symmetry surfaces: $(100)$ bent along $[011]$,
and $(111)$.

We start from the elastic stiffnesses $C_{ij}$ which are related to the
compliances $\sigma _{ij}$ by the equations \cite{kittel}:
\begin{eqnarray}
\sigma _{11} &=&\frac{C_{11}+C_{12}}{\left( C_{11}-C_{12}\right) \left(
C_{11}+2\,C_{12}\right) } \\
\sigma _{12} &=&-\frac{C_{12}}{\left( C_{11}-C_{12}\right) \,\left(
C_{11}+2\,C_{12}\right) }  \nonumber \\
\sigma _{44} &=&\frac 1{C_{44}}.  \nonumber
\end{eqnarray}
If the uniaxial bending is not along a $[100]$ direction and for surface
orientations other than $\left( 100\right) $, the compliances $\sigma _{11}$
and $\sigma _{12}$ which appear in (\ref{eq:Young}) have to be replaced by
effective elastic constants, which in the $(100)$ case, with bending along a 
$[011]$ direction, have the form:

\begin{eqnarray}
\sigma _{11}^{^{\prime }} &=&\sigma _{11}-\frac 12\left( \,\sigma
_{11}+\sigma _{12}+\frac 12\sigma _{44}\right) \label{eq:prime}\\
\sigma _{12}^{^{\prime }} &=&\sigma _{12}+\frac 12\left( \,\sigma
_{11}+\sigma _{12}+\frac 12\sigma _{44}\right) ,  \nonumber
\end{eqnarray}
whereas in the $\left( 111\right) $ case have the form: 
\begin{eqnarray}
\sigma _{11}^{^{\prime }} &=&\sigma _{11}-\frac 12\left( \,\sigma
_{11}+\sigma _{12}+\frac 12\sigma _{44}\right) \\
\sigma _{12}^{^{\prime }} &=&\sigma _{12}+\frac 16\left( \,\sigma
_{11}+\sigma _{12}+\frac 12\sigma _{44}\right) .  \nonumber
\end{eqnarray}

\section{Application to metal surfaces}

\label{sec4}

\subsection{Implementation}

The first goal of the variable-curvature MD simulation is to obtain an
equilibrium value for the curvature in order to extract the surface stress
difference, according to equation (\ref{eq:stoney}). We can then compare the
outcome with the surface stress difference calculated independently using
Kirwood's formula \cite{kirkwood}, and assess the success of the method.
We will also verify the validity of the approach by analyzing the
behavior of the curvature as a function of the slab thickness.

We will present two different exemplifications. Far from being exhaustive,
these results are only meant to show the feasibility of the method in view
of later applications to surface physics.

First we simulated several Au slabs with (100) orientation and
different types of reconstruction on the
two sides. This test can show the sensitivity of the method:
differences of the order of $5$ $\rm meV/\AA ^2$ can be appreciated.
Next we examined the effect of isotropic or anisotropic
adsorbates on the surface stress, as exemplified by Pb/Au(100).

All simulations were performed using many-body potentials of the glue type 
\cite{ercpartos}. They have the general form:
\begin{eqnarray}
V &=&\frac 12\sum_{i,\,\,j}\Phi \,\left( \,r_{\,ij}\right) +\sum_iU\,\left(
n_i\right)   \label{glue} \\
n_i &=&\sum_j\rho \,\left( \,r_{ij}\right) ,  \nonumber
\end{eqnarray}
where $n_i$ is a generalized atomic coordination. $\Phi (r),$ $U\,(n)$ and $%
\rho \,\left( r\right) $ are empirically constructed, by fitting several
properties of the system. Well tested glue potentials are available 
\cite{webglue} for Au \cite {ercpartos} and Pb \cite{lead}.
This scheme can be extended to binary systems 
by introducing a mixed two-body potential $\Phi _{AB}(r)$, and
assuming $n_i$ to be a linear superposition of density contributions
$\rho_A$ and $\rho_B$ supplied respectively by $A$- and $B$-type atoms 
to site $i$ \cite{orio}:
\begin{eqnarray}
V &=&\frac 12\left[ \sum_{i\in A}\left( \sum_{j\in A}\Phi _{AA}\left(
\,r_{ij}\right) +\sum_{j\in B}\Phi _{AB}\left( \,r_{ij}\right) \right)
+\sum_{i\in B}\left( \sum_{j\in A}\Phi _{AB}\left( \,r_{ij}\right)
+\sum_{j\in B}\Phi _{BB}\left( \,r_{ij}\right) \right) \right]   \nonumber \\
&&+\sum_{i\in A}U_A\,\left( n_i\right)
  +\sum_{i\in B}U_B\,\left( n_i\right) ,  
\label{eq:multiglue}
\end{eqnarray}
with
\begin{equation}
n_i=\sum_{j\in A}W_A\rho _A\left( r_{ij}\right) +\sum_{j\in B}W_B\rho
_B\,\left( r_{ij}\right) .
\end{equation}
where $W_A$ and $W_B$ are suitable weights. 
In our case $W_{\rm Au}=1.045$ and $W_{\rm Pb}=0.957$.

\subsection{A clean surface: change of surface stress with reconstruction}

\label{sub:results1}

We prepared a family of Au(100) slabs composed by about 600 atoms per layer 
($L_x=28.78\,\rm\AA$, $L_y=172.69\,\rm\AA$),
and different thicknesses of 8, 12, 16 and 20 layers,
with [011] as the bending direction.

In order to extract numerical values for the surface stress from equation (%
\ref{eq:stoney}), we need the correct elastic constants for this
case. For the case of Au (as described by the glue model), the elastic 
stiffnesses have been calculated \cite{ercpartos}, 
giving 
\begin{eqnarray}
C_{11} &=&2.203\cdot 10^7\,{\rm N/cm}^2 \\
C_{12} &=&1.603\cdot 10^7\,{\rm N/cm}^2  \nonumber \\
C_{44} &=&0.600\cdot 10^7\,{\rm N/cm}^2.  \nonumber
\end{eqnarray}
With these values, using (\ref{eq:Young}) and (\ref{eq:prime}), 
the Young's modulus and the Poisson's ratio for Au$\left( 100\right) ,$ with
bending along [011], are $Y_{(100)}=1.322\cdot 10^7\,{\rm N/cm}^2$ 
and $\nu _{\left( 100\right) } =0.1022$. 

Both $(100)$ surfaces were prepared in the reconstructed state,
characterized by a close-packed triangular overlayer on a square substrate.
We chose a fixed reconstructed $(1\,{\rm \times }\,5)$ structure (six $[011]$
rows on top of five) for the reference (lower) surface of the plate, and we
changed the nature of the upper surface, ranging from $(1\,{\rm \times }$%
\thinspace $3)$ (\thinspace {\it i.e. }four surface rows over three
substrate rows, a ``compressive'' situation) to a $(1\,{\rm \times \,}20)$
structure, (a ``tensile'' situation), touching the $(1{\rm \ \times }${\rm \ 
}$4),$ $(1\,{\rm \times \,}5),$ $(1\,{\rm \times \,}6),$ $(1\,{\rm \times \,}%
10),$ $(1\,{\rm \times \,}12),$ and $(1\,{\rm \times \,}15)$ structures. In
the $(1\,{\rm \times }$\thinspace $5)$ case no bending was expected or
observed, since the surfaces on the two sides were identical. We also considered 
$(11\,{\rm \times \,}3),$ $(11\,{\rm \times \,}4),$ $(11\,{\rm \times \,}%
5), ...$ obtained by adding $[01\bar{1}]$ rows. The lowest energy surface in
the glue model is $(34\,{\rm \times \,}5)$\cite{ept}, in the
experiment is close to $(28$\thinspace ${\rm \times \,}5),$ but all 
$({M\,\times
\,}5)$ surfaces differ very slightly in free energy \cite
{yamazaki,leed_Au100,prof_Au100,STM_Au100}. The essential point here will be
that surfaces which differ only very little in energy, for example $(1\,{\rm %
\times \,}5)$ $(102.3\,{\rm meV/\AA }^2)$ and $(1\,{\rm \times \,}12)$ $%
(103.2$ ${\rm meV/\AA }^2)$ may differ enormously in surface stress.

A typical snapshot of the simulation is shown in Figure 2.

Figure 3 shows the time evolution of the curvature $k$.
Starting from zero, the slab curvature reaches the equilibrium
value very rapidly. A damping term, although not really necessary, has been
added in order to speed up the convergence.

We verified that, as it should be and as can be seen from the figure, the
value of the mass $W$ is irrelevant in determining the equilibrium value of $%
k.$ In the $(1{\rm \,\times \,}3)$ and $\ (1{\rm \ \times }${\rm \ }$4)$
cases, the equilibrium $k$ was positive, whereas in the other cases it was
negative. 
The simulation was performed
at low temperature (from a few degrees to $300\,$K). The difference in surface
stress between upper and lower surface has been extracted using (\ref
{eq:stoney}). The surface stress of the two isolated surfaces has been
independently obtained from MD\ forces using the standard Kirkwood-Buff formula 
\cite{kirkwood}. The comparison of the stress differences is excellent, as can
be seen from Figure 4,
showing that the variable curvature method works. The only deviation is for $%
(1{\rm \,\times \,}3)$, and clearly attributable to excessive curvature.

The calculation is repeated with a variety of different unit cells in the
top surface; by adding a $[01\bar{1}]$ row every ten, we obtain $(11\,{\rm %
\times \,}3),$ $(11\,{\rm \times \,}4),$ $(11\,{\rm \times \,}5),...$ unit
cells instead of $(1{\rm \ \times }${\rm \ }$3),$ $(1{\rm \ \times }${\rm \ }%
$4),$ $(1\,{\rm \times \,}5),...$ The lower $(1\,{\rm \times \,}5)$ surface
of the slab was left unchanged.
The stress variations are cleary very large, both relative to the basic
stress of $(1\,{\rm \times \,}5)$ $(203.12$\thinspace {\rm meV/\AA }$^2)$
and to the surface energy $(\sim 100\,${\rm meV/\AA }$^2).$ We
judge that the smallest surface stress difference detectable with our kind
of simulation is of the order of $10\,${\rm meV/\AA }$^2.$ This sensitivity
should be very important in the study of phase transitions.

We have also verified the linear dependence of the curvature from the
inverse square of the thickness predicted by equation (\ref{eq:stoney}).
Figure 5 shows the fit with 8, 12, 16 and 20 layers, and the agreement is
rather good.

\subsection{Change of surface stress with adsorption}

\label{sub:results2}

The surface stress is obviously a strong function of adsorption. As an
example, we will consider here Pb on Au. Underpotential Deposition (UPD) 
\cite{UPD} has been recently used together with STM\ techniques to exploit
the formation of 2D phases of Pb on a substrate of Ag(100) and Au(100).
Through different electrochemical potentials, one can obtain different
phases, such as Au(100)-${\rm c}(2\,{\rm \times \,}2)$ Pb, Au(100) ${\rm c}%
\left( 3\sqrt{2}\,{\rm \times \,}\sqrt{2}\right) {\rm R\,}45^{\circ }{\rm \ }
$Pb, Au(100) {\rm c}$\left( 6\,{\rm \times \,}2\right) $ Pb. In particular,
a close packed layer of lead shows a slight contraction in both $[011]$
directions of the quadratic substrate lattice; the final structure is pinned
to the substrate lattice at equivalent adsorption sites, leading to an
Au(100)-{\rm c}$\left( 6\,{\rm \times \,}2\right) $ moir\'{e} superstructure.

Using the potential in eq.\ (\ref{eq:multiglue}), we found precisely this 
{\rm c}$\left( 6\,{\rm \times \,}2\right) $ superstructure as a local energy
minimum, through quenching of a sample prepared with a perfect substrate of
Au(100), and a triangular overlayer of Pb with the same density
of a bulk (111) layer.
In the optimized structure, the $(6\,{\rm \times
\,}2)$ pattern is evident (Figure 6(a)).
Domain walls appear as the result of overlayer contraction.
We were able to obtain a perfect $(6\,{\rm \times \,}2)$ (Figure 6(b))
by starting from an initial configuration with a denser overlayer.
Deeper energy minima are obtained by the simple commensurate square
overlayers with vacancies as in the case shown in Figure 6(c).

We performed variable curvature simulations on both the imperfect 
and the perfect {\rm c}$\left( 6\,{\rm \times \,}2\right) $, 
and on the square overlayer with vacancies,
with Au(100) $(1\,{\rm \times \,}4)$ as the reference 
($\tau _{22}=147.52$ {\rm meV/\AA }$^2).$
For {\rm c}$\left( 6\,{\rm \times \,}2\right)$, which is anisotropic,
two different simulations were carried out in order to access separately $%
\tau _{11}$ and $\tau _{22}$ (the adlayer was rotated of 90 degrees, 
so as to keep $\tau_{22}$ as reference).

For the $(6\,{\rm \times \,}2)$ structure with domain walls we found $\tau
_{22}=88.10$ {\rm meV/\AA }$^2$ ($\tau _{11}$ was not measured in this case),
and for the perfect $(6\,{\rm \times \,}2)$\ structure 
$\tau_{11}=72.41$ {\rm meV/\AA }$^2$ and $\tau _{22}=38.52$ {\rm meV/\AA }$^2.$
For the square structure with vacancies, we found 
$\tau_{22}=40.34\,{\rm meV/\AA}^2$.
The stress anisotropy is therefore large, and the effect of the domain walls 
is also remarkable.

\section{Conclusions}

\label{sec5} In this work we have introduced a variable curvature slab
simulation method which represents a promising technique to study properties
of crystal surfaces, in particular surface stress differences.

The first tests of the method, based on the comparison with known results,
reveal a good accuracy, in the range of $10$ ${\rm meV/\AA }^2,$ and easy
applicability. We are now considering studies of a variety of phenomena,
including surface phase transitions, using this approach. 

\section*{Acknowledgments}

We would like to thank Francesco Di Tolla for useful discussions
and help.   We acknowledge support from EU through 
contracts ERBCHBGCT920180, ERBCHBGCT940636 and ERBCHRXCT930342,
and from INFM through PRA LOTUS.

\section*{Appendix A. Interpretation of the kinetic energy of the
curvature.}

If the arc $s_3=0$ is forced to have length $L_y$ during the whole
simulation, the volume $\Omega $ of the sample is guaranteed to remain
constant and equal to $L_xL_yL_z.$ This is easily seen by performing the
integral in polar coordinates.

The difference in surface area between upper and lower layer is of course:
\begin{equation}
\Delta A=L_x\left[ \theta _{MAX}\,\left( R+\frac{L_z}2\right) -\theta
_{MAX}\,\left( R-\frac{L_z}2\right) \right] =L_xL_z\theta
_{MAX}=L_xL_yL_z\,k=\Omega \,k.
\end{equation}
The relative time variation of this area is
\begin{equation}
\frac{{\rm d}}{{\rm dt}}\left( \frac{\Delta A}{\bar{A}}\right) =\frac V{%
L_xL_y}\frac{{\rm d}}{{\rm dt}}k=L_z\dot{k}
\end{equation}
Therefore, the fictitious kinetic energy term is:
\begin{equation}
{\em \tilde{T}}=\frac 12W\,\dot{k}^2=\frac 12\left( \frac W{L_z^2}\right)
\left( \frac{{\rm d}}{{\rm dt}}\left( \frac{\Delta A}{\bar{A}}\right)
\right) ^2=\frac 12\left( \frac W{\Omega ^2}\right) \left( \frac{{\rm d}}{%
{\rm dt}}\left( \Delta A\right) \right) ^2.
\end{equation}

This result is simple but important because allows a connection with
Andersen molecular dynamics \cite{andersen}:
whereas in Andersen's formulation the extra degree of freedom is the volume 
$\Omega $ and the term $\frac 12W\,\dot{\Omega}^2$ is the kinetic energy
tied to volume variations, in this case there is a variation in the area
difference between the two faces of the sample. The analogy could be
brought further, adding to our Lagrangian a term $-P\,k$, which is the
reversible work of a radial force decaying as the square of the curvature
radius, acting identically on all particles, whereas the term $-P\,\Omega $
in Andersen's formulation is the reversible work of the hydrostatic pressure
during the variation of the cell volume.
Such an external force, acting as a torque,
could be used to force the equilibrium curvature
to a different value from the one suggested by stress imbalance between the
two surfaces. Applications of this point are presently under study.

\section*{Appendix B. Constrained dynamics for fixing the center of mass}

\subsection{The velocity Verlet as a predictor-corrector.}

Following \cite{allen}, it is possible to cast a particular velocity Verlet
algorithm as a predictor-corrector algorithm.

One can start from
\begin{eqnarray}
{\bf r}_0(t+\delta t) &=&{\bf r}_0\left( t\right) +{\bf r}_1\left( t\right) +%
{\bf r}_2\left( t\right) \\
{\bf r}_1(t+\delta t) &=&{\bf r}_1\left( t\right) +{\bf r}_2\left( t\right) +%
{\bf r}_2\left( t+\delta t\right)  \nonumber
\end{eqnarray}
where, as usual, 
\begin{equation}
{\bf r}_n\left( t\right) =\frac 1{n!}\left( \delta t\right) ^n\frac{{\rm d}^n%
{\bf r}_0}{{\rm d}t^n}.  \label{eq:asusual}
\end{equation}
This can be written as a two stages predictor-corrector,
with a force evaluation in between:
\begin{eqnarray}
{\bf r}_0(t+\delta t/2) &=&{\bf r}_0\left( t\right) +{\bf r}_1\left(
t\right) +{\bf r}_2\left( t\right) \\
{\bf r}_1(t+\delta t/2) &=&{\bf r}_1\left( t\right) +{\bf r}_2\left(
t\right)  \nonumber \\
{\bf r}_0(t+\delta t) &=&{\bf r}_0(t+\delta t/2)  \nonumber \\
{\bf r}_1(t+\delta t) &=&{\bf r}_1(t+\delta t/2)+{\bf r}_2\left(
t+\delta t\right)   \nonumber
\end{eqnarray}
where ${\bf r}_2\left( t+\delta t\right) $ is obtained
by the equations of motion. 
In this case the predicted values for the positions coincide with the corrected
ones, whereas the velocities are corrected (second-order, or three values,
predictor-corrector).

When constraints are present, the forces acting on the $i$-th particle can
be written as:
\begin{equation}
{\bf F}_i={\bf f}_a+{\bf g}_a\approx {\bf f}_a+{\bf g}_a^{(x)}.
\end{equation}
The constraints forces ${\bf g}_a$ at each stage are approximated with ${\bf %
g}_a^{(x)}$, where $(x)=r,v$,
so as to ensure that the positions and the velocities, respectively,
satisfy the constraints, and the
quantities at the various stages have to be corrected 
to keep into account the constraints forces.

\subsection{The bent plate case.}

In the case of varying curvature molecular dynamics, the bending leads to a
drift towards the center of curvature of the center of mass. 
In order to fix the $s_3$ component of 
the center of mass in the origin, the only constraint to
be satisfied is (assuming all the particles to have the
same mass for sake of simplicity)
\begin{equation}
\sigma \equiv \sum\limits_{i=1}^Ns_{3i}=0
\end{equation}
where $s_{3i}$ are the scaled coordinates in our curvilinear system (in
this appendix, $s,\dot{s}$ and $\ddot{s}$ are normalized according to (%
\ref{eq:asusual})).
The time derivative of the constraint equation gives a constraint on
velocities:
\begin{equation}
\dot{\sigma}\equiv \frac 1{\delta t}\sum\limits_{i=1}^N\dot{s}_{3i}=0.
\end{equation}
The Lagrangian must then be completed with a term 
$\lambda_R\,\sum\limits_{i=1}^Ns_{3i}
+\lambda_V\,\sum\limits_{i=1}^N\dot{s}_{3i}$.

Two different approximations to the constraint forces have to be chosen so that
positions and velocities satisfy the constraints exactly. In this 
particular case the forces turn out to be simple constants:
\begin{eqnarray}
g_{3i}^{(r)} &=&-\lambda _R \\
g_{3i}^{(v)} &=&-\lambda _V  \nonumber
\end{eqnarray}
The first equation, involving positions, is: 
\begin{equation}
s_{3i}(t+\delta t)=s_{3i}\left( t\right) +\dot{s}_{3i}\left( t\right) +\ddot{%
s}_{3i}\left( t\right) -\frac{\left( \delta t\right) ^2}{2m_i}\lambda _R
\end{equation}
$\lambda _R$ has to be chosen in order to satisfy the position constraint at
time $t+\delta t.$:
\begin{equation}
\begin{tabular}{c}
$\sum_{i=1}^N\left( s_{3i}\left( t\right) +\dot{s}_{3i}\left( t\right) +%
\ddot{s}_{3i}\left( t\right) -\frac{\left( \delta t\right) ^2}{2m_i}\lambda
_R\right) =0$ \\ 
$\lambda _R=2m_i\frac{\sum_{i=1}^N\left( s_{3i}\left( t\right) +\dot{s}%
_{3i}\left( t\right) +\ddot{s}_{3i}\left( t\right) \right) }{N\left( \delta
t\right) ^2}.$%
\end{tabular}
\end{equation}
Therefore the former equation becomes:
\begin{equation}
s_{3i}(t+\delta t)=s_{3i}\left( t\right) +\dot{s}_{3i}\left( t\right) +\ddot{%
s}_{3i}\left( t\right) -\frac{\sum_{j=1}^N\left( s_{3j}\left( t\right) +\dot{%
s}_{3j}\left( t\right) +\ddot{s}_{3j}\left( t\right) \right) }N.
\end{equation}

The second equation, involving velocities, is:
\begin{equation}
\dot{s}_{3i}(t+\delta t)=\dot{s}_{3i}\left( t\right) +\ddot{s}_{3i}\left(
t\right) -\frac{\left( \delta t\right) ^2}{2m_i}\lambda _R+\ddot{s}%
_{3i}\left( t+\delta t\right) -\frac{\left( \delta t\right) ^2}{2m_i}\lambda
_V.
\end{equation}
$\lambda _V$ must satisfy
\begin{equation}
\begin{tabular}{c}
$\sum_{i=1}^N\left( \dot{s}_{3i}\left( t\right) +\ddot{s}_{3i}\left(
t\right) -\frac{\sum_{j=1}^N\left( s_{3j}\left( t\right) +\dot{s}_{3j}\left(
t\right) +\ddot{s}_{3j}\left( t\right) \right) }N+\ddot{s}_{3i}\left(
t+\delta t\right) -\frac{\left( \delta t\right) ^2}{2m_i}\lambda _V\right)
=0 $ \\ 
$\lambda _V=2m_i\frac{\sum_{i=1}^N\left( \ddot{s}_{3i}\left( t+\delta
t\right) -s_{3i}\left( t\right) \right) }{N\left( \delta t\right) ^2}.$%
\end{tabular}
\end{equation}
and the equation becomes
\begin{eqnarray}
\dot{s}_{3i}(t+\delta t) &=&\dot{s}_{3i}\left( t\right) +\ddot{s}_{3i}\left(
t\right) -\frac{\sum_{i=1}^N\left( s_{3i}\left( t\right) +\dot{s}_{3i}\left(
t\right) +\ddot{s}_{3i}\left( t\right) \right) }N \\
&&+\ddot{s}_{3i}\left( t+\delta t\right) -\frac{\sum_{j=1}^N\left( \ddot{s}%
_{3j}\left( t+\delta t\right) -s_{3j}\left( t\right) \right) }N.  \nonumber
\end{eqnarray}

\newpage

\section*{Figure captions}

\begin{itemize}
\item  Figure 1. The geometry of a bent plate with curvilinear coordinates; $%
s_1$ and $s_2$ lie in the neutral plane (shown as lighter in the figure)
whereas $s_3$ is vertical. $R$ is the curvature radius measured at $s_3=0,$
and $\theta _M$ is the bending angle.

\item  Figure 2. A typical snapshot of the simulation, with a positive value
of the curvature. The top surface is a $(1\,{\rm \times }\,10)$ and the
lower surface a $(1\,{\rm \times }\,5)$, both hexagonally reconstructed
Au(100). The corrugations of both surfaces are evident. This particular
sample will invert its curvature during the simulation, and the equilibrium
value of $k$ will become negative.

\item  Figure 3. Time evolution of the curvature $k$ for two different
values of the `mass' $W.$ On the right axis, the bending angle $\theta _M$
is shown.

\item  Figure 4. Comparison between variable curvature simulations and
Kirkwood-Buff formula, with different structures at the upper surface of the
sample. The lower surface is a $(1\,{\rm \times }\,5)$ in every case. 
Right ordinate axis: equilibrium value for the curvature. 
Left ordinate axis: corresponding
difference in surface stress, according to Stoney's equation. In order to
obtain the absolute value of the surface stress component $\tau _{22}$ of
the upper surface, the reference $(1\,{\rm \times }\,5)$ result $(203.12$ 
{\rm meV/\AA }$^2)$ has to be added.

\item  Figure 5. Dependence of the curvature upon thickness in the case of $%
(1\,{\rm \times }\,10)/(1\,{\rm \times }\,5)$. The inverse square
dependence predicted by Stoney's equation is recovered, and the stress
difference is extracted from the fit.

\item  Figure 6. Different structures for the Pb overlayer (grey and
black atoms) over a square
Au(100) substrate (white atoms). (a): $(6\,{\rm \times \,}2)$ structure with domain walls.
Black atoms are higher, two MD boxes are shown along $x$; (b): $(6\,{\rm %
\times \,}2)$ without domain walls; (c) square adlayer with vacancies.
\end{itemize}


\begin{references}
\bibitem{extens}  The extension to potentials with angular dependence, such
as Tersoff potentials, requires of course the derivation of more cumbersome
equations of motion, but is in principle possible.
\end{references}
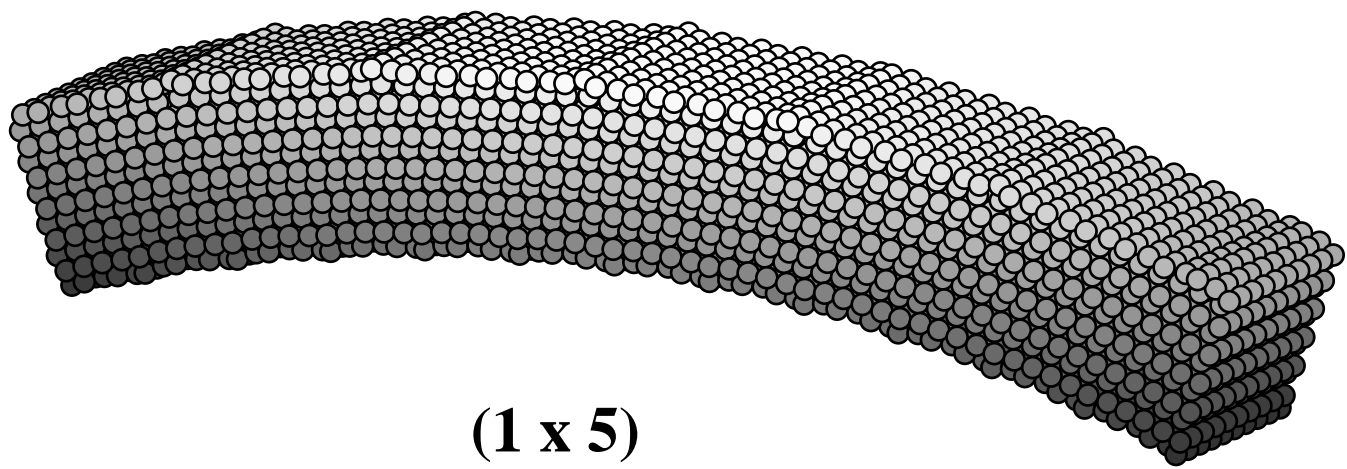
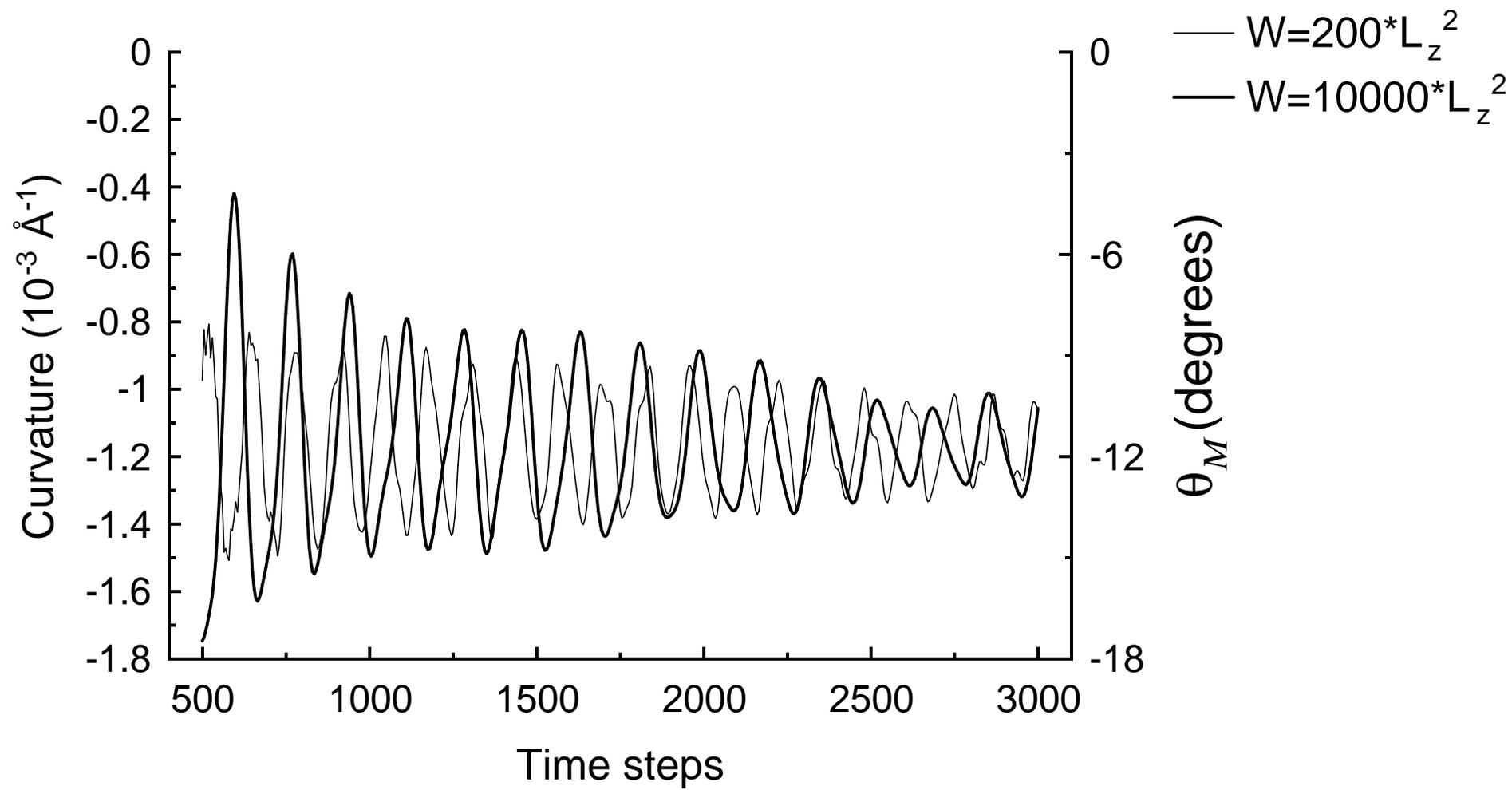
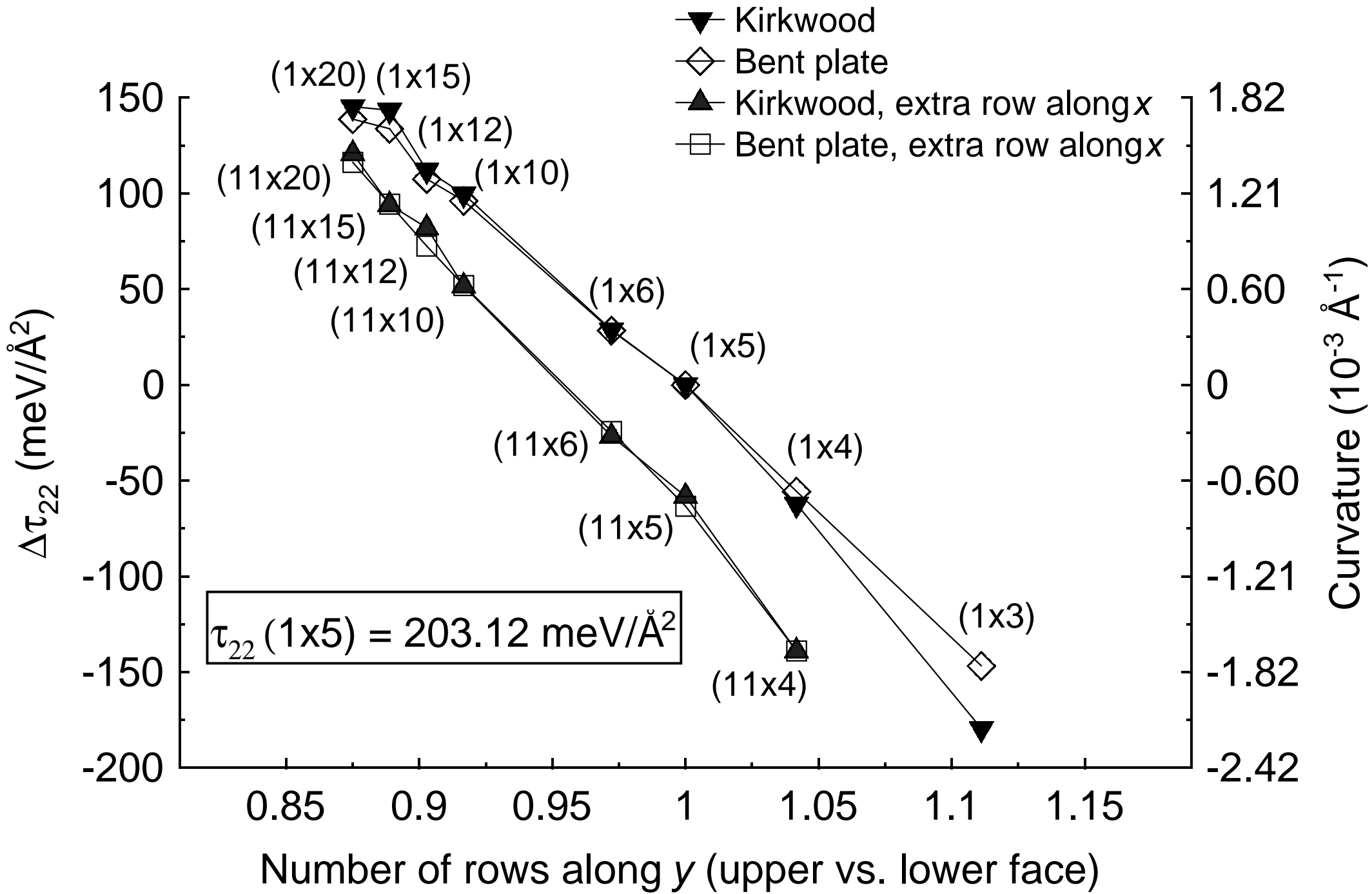
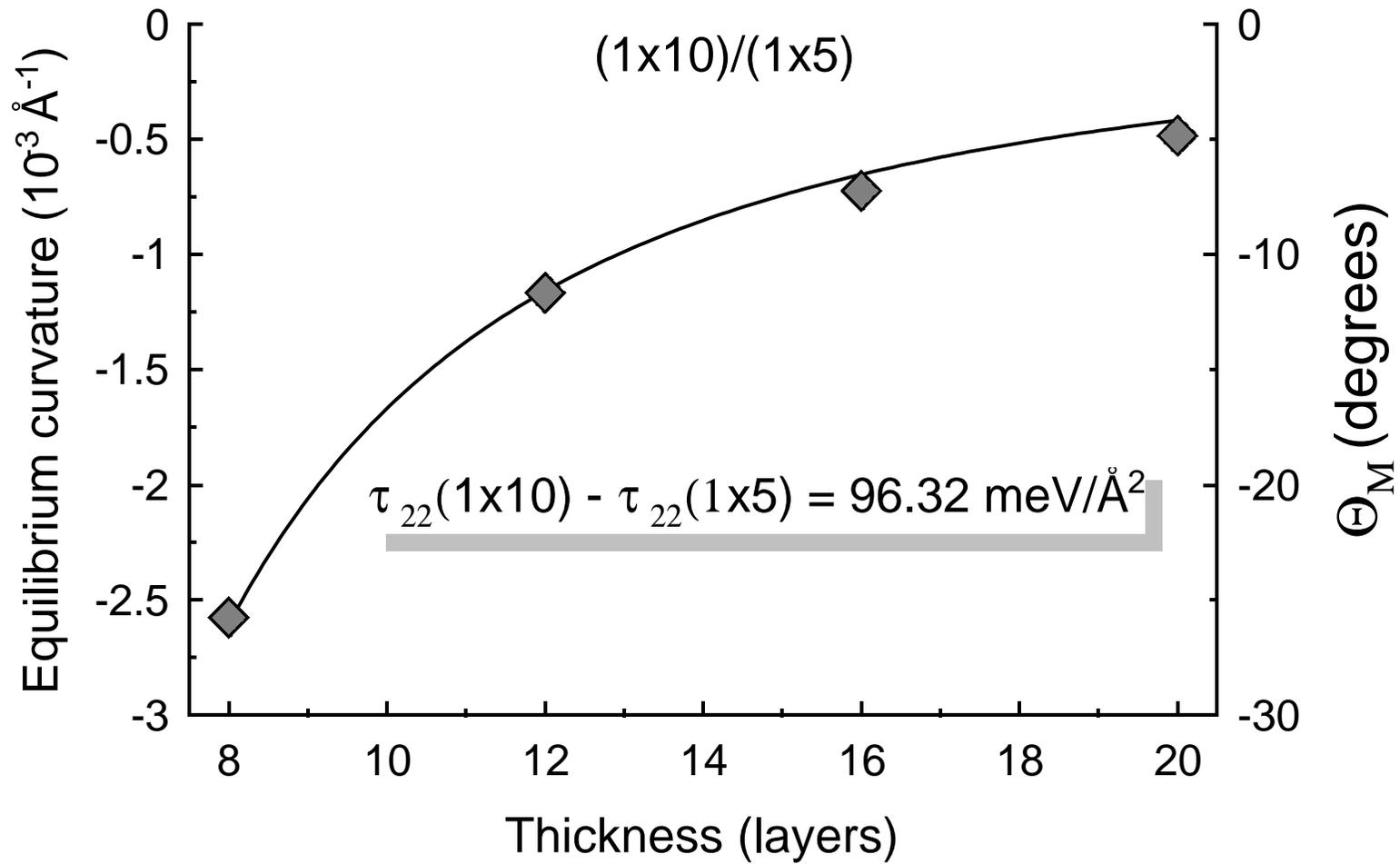
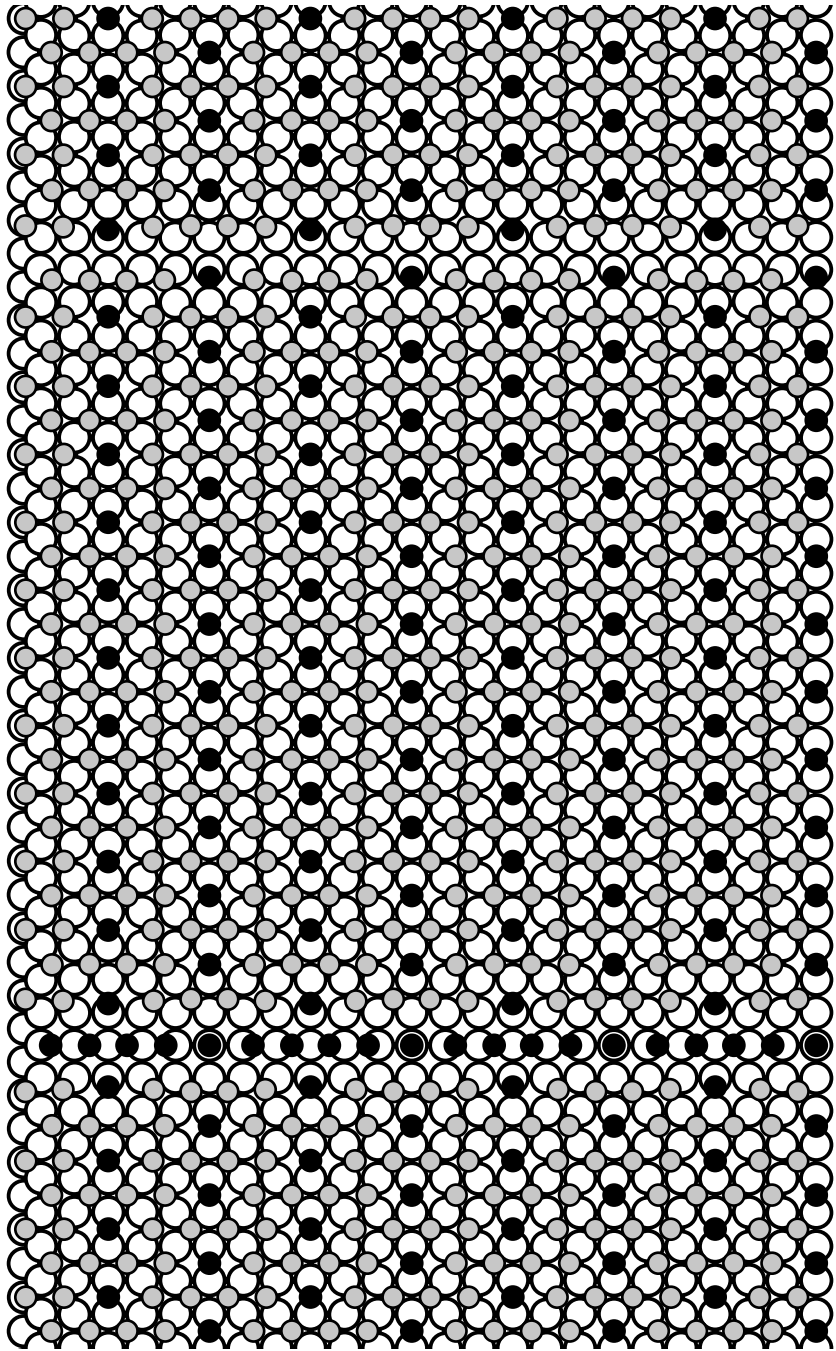
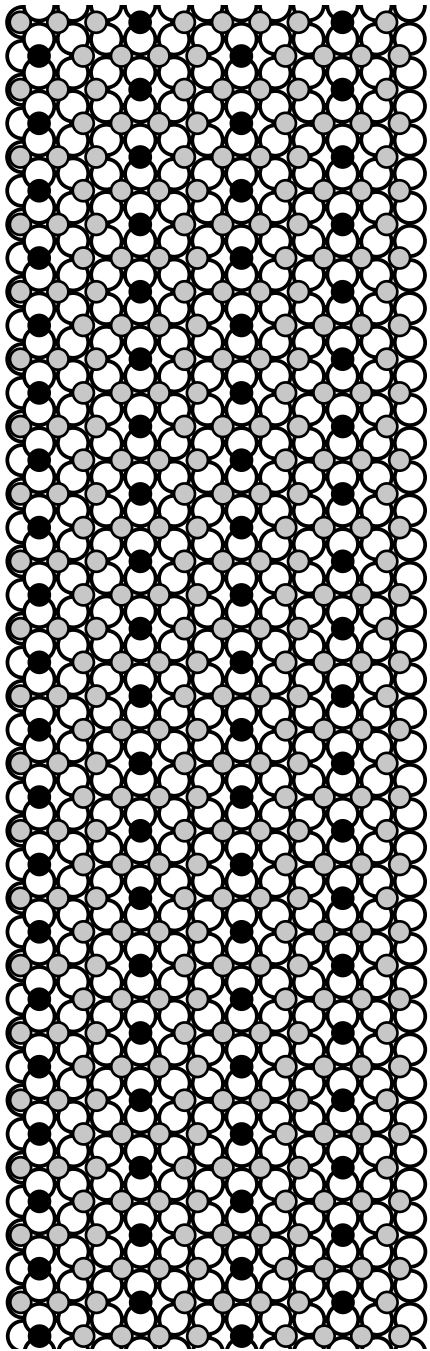
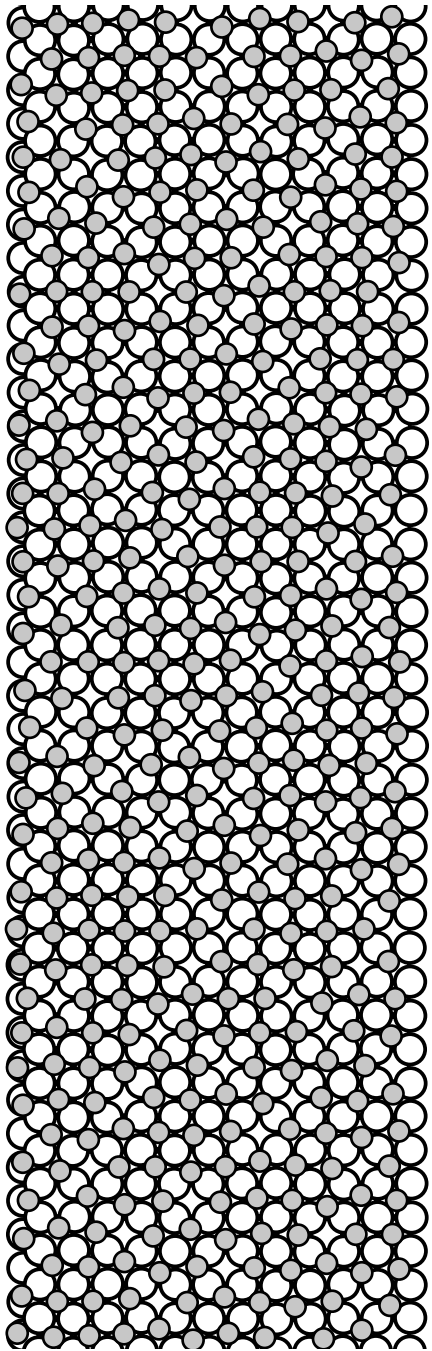
\end{document}